\documentstyle[11pt]{article}

\def\slashchar#1{\setbox0=\hbox{$#1$}           
   \dimen0=\wd0                                 
   \setbox1=\hbox{/} \dimen1=\wd1               
   \ifdim\dimen0>\dimen1                        
      \rlap{\hbox to \dimen0{\hfil/\hfil}}      
      #1                                        
   \else                                        
      \rlap{\hbox to \dimen1{\hfil$#1$\hfil}}   
      /                                         
   \fi}
\begin{document}
\begin{titlepage}
\begin{center}
May, 1998      \hfill     BUHEP-98-13\\
\vskip 0.2 in
{\Large \bf Effective Field Theory for a Three-Brane Universe}

\vskip .2 in
         {\bf   Raman Sundrum\footnote{email: sundrum@budoe.bu.edu.}}
        \vskip 0.3 cm
       {\it Department of Physics \\
Boston University \\
Boston, MA 02215, USA}
 \vskip 0.7 cm 

\begin{abstract}
A general effective field theory formalism is presented which 
describes the low-energy dynamics of a 3-brane universe. In 
this scenario an arbitrary four-dimensional particle theory, 
such as the Standard Model, is constrained to live on the 
world-volume of a (3+1)-dimensional hypersurface, or ``3-brane'', 
which in turn fluctuates in a higher-dimensional, gravitating 
spacetime. The inclusion of chiral fermions on the 3-brane is 
given careful treatment. The power-counting needed to renormalize 
quantum amplitudes of the effective theory is also discussed. The 
effective theory has a finite domain of validity, restricting it
to processes at low enough energies that the internal structure 
of the 3-brane cannot be resolved. 
\end{abstract}
\end{center}

\end{titlepage}

\section{Introduction}

How many dimensions do we live in? 
 Macroscopically, we {\it see} and {\it feel} three
spatial dimensions using electromagnetism. Furthermore,
 Newton's $1/r^2$ Law of
Gravity follows from general relativistic principles in $3 + 1$
dimensions. Microscopically, calculations based on
$(3+1)$-dimensional spacetime are in excellent accord
 with our most sensitive experimental tests of the Standard Model (SM). 
Yet, it is well known that extra spatial dimensions are possible if
they are compactified at sufficiently small radii. To resolve a compact
dimension, it must be probed by quanta with wavelengths smaller than its
radius. Presently, our sharpest SM probes have wavelengths as short as
$ \sim 10^{-16}$ cm. Apparently this provides the upper bound on the
radii of any higher dimensions occurring in nature. But this conclusion
is based on the assumption that {\it all} particle species move in
the same number of dimensions as the SM.  This assumption is
implicit in the standard Kaluza-Klein approach to higher dimensions, which,
until recently, played a central role in  string theory. 

Can non-SM particles see extra dimensions that the SM does not? At
present, we know of only  one non-SM state, the
graviton, but 
others might exist if they are weakly-coupled  or massive enough 
to have escaped detection thusfar. Let us suppose there are extra spatial
dimensions in which gravity can propagate\footnote{Indeed since gravity
  is so intimately tied to spacetime, it would be hard to conceive of
  gravity {\it not} being present in all the extra dimensions.} but
in which the SM cannot. We can think of the SM as being ``stuck'' at
some definite  position in the extra dimensions. That is, if the full
``bulk'' spacetime is really $d$-dimensional, $d > 4$, then we are considering
the SM to be confined to a $(3+1)$-dimensional hypersurface. Since the bulk
spacetime contains gravity and is therefore dynamical, the hypersurface
cannot be rigid, but must also be dynamical. We can borrow some
string-theory parlance and call this dynamical hypersurface  a
``3-brane''.\footnote{In string theory, ``3-brane'' has a more
  restricted usage.} The basic idea that the standard particles are
confined to a 3-brane in higher dimensions goes back at least to ref. 
\cite{rubakov}. Very recently, it has been discussed in refs. \cite{1} 
and \cite{2}, as a possible means of addressing the unnatural
hierarchy between the weak and Planck scales. These papers give an
up-to-date analysis of some of the theoretical and phenomenological
possibilities. Related ideas involving 3-brane universes and/or
relatively low compactification scales appear in refs. [4 -- 14].

Let us now impose the constraint that
Newton's $1/r^2$ Law, or more generally, four-dimensional
general relativity, is experimentally verified at macroscopic
distances. Therefore by these distances, gravity must also be confined
to $3 + 1$ dimensions, the most obvious way being by compactifying the
extra dimensions. But now the compactification radii are not constrained
to be smaller than $10^{-16}$ cm because SM particles are not direct
probes of the higher dimensions. To
resolve the higher dimensions we must use gravity which has only
been tested down to a distance of a centimeter. The radii of the 
extra dimensions need only be smaller than this for us to not yet have
observed them. An exciting possibility is that future short-distance tests
of gravity may do so \cite{1} \cite{price}.

Even if the compactification radii are too small to be seen directly in
the forseeable future, they can lead to interesting indirect
effects. One intriguing possibility is provided by the work initiated in
refs. \cite{horava} in the context of M-theory, literally a case of parallel
universes. We can effectively have two 3-branes,
separated in a $(4+1)$-dimensional bulk spacetime. A supersymmetric and gauge
extension of the SM lives on one 3-brane, while a strongly-interacting
supersymmetric  
{\it hidden sector} lives on the other. The hidden sector dynamics can
trigger  spontaneous supersymmetry breaking on its 3-brane, which is 
transmitted by  bulk modes  to the SM 3-brane, where it shows up
 as soft supersymmetry breaking. (For an interesting study of this
 phenomenon in a simple setting see ref. \cite{peskin}.)

If we do live on a 3-brane, what is it made of? 
It has been understood for some time
that a  quantum field theory can contain 
topological defects of various types and dimensionality, which can have
low-energy particle-like modes trapped on them. It therefore seems
plausible that our 3-brane is a $(3+1)$-dimensional defect in a higher
dimensional field theory, and the SM particles are some of the light
modes trapped on the defect. This is the scenario advocated in ref. \cite{1}.
While it is known how to build theories of
this type where scalars and fermions live on the defect, it is
still problematic to  obtain  low-energy four-dimensional 
 vector (gauge) fields, although
some new ideas are  pursued in refs. \cite{dvali} 
\cite{1}. It is therefore fair
to say that while the 3-brane scenario has
reasonable support within quantum field theory, 
 we are still some way from a realistic model.

The situation improves if we consider superstring theory, the direction
taken in refs. \cite{2} \cite{tye}. Intrinsic to
the theory are Dirichlet-branes (D-branes), defects of varying
dimensionality on which open strings can end. See ref. \cite{polchinski}
for a review.  The short open string
modes trapped on D-branes can include light gauge fields, fermions and 
scalars, all the basic ingredients for realistic theories. It is
possible to construct a variety of sandwiches of D-branes and
strings (reveiwed in ref. \cite{giveon}), 
which reduce at low energies to  interesting
particle theories effectively living in four dimensions, weakly
coupled to gravity and other modes which propagate in higher
dimensions.\footnote{The way effective 3-branes arise in the scenario of
  ref. \cite{horava}, mentioned above, is more subtle. 
See ref. \cite{ovrut} for a discussion.} However, a realistic SM sector  
has not yet been engineered in this way.

The purpose of the present paper is to show how one can construct
realistic 
{\it effective field theories} to study  the low-energy
consequences 
of the 3-brane scenario, in a systematic, economical and elegant way.  The
only degrees of freedom that appear are those that matter at low
energies, the purely high-energy degrees of freedom are considered to be
integrated out.   Even if a
fundamental description were known, the most efficient  way of pursuing the
low-energy dynamics would be to {\it match} the fundamental theory to
such an effective field theory,
 and then use the latter for calculations and insight.

Roughly speaking, the low-energy domain 
restricts us to processes which cannot resolve the stucture of
the 3-brane. In many ways this approach is analogous to the chiral
lagrangian approach to  pion dynamics, where the pion is
treated as a point-particle whose internal quark-glue structure is outside the
low-energy domain of validity. 
We will simply {\it assume} that there is {\it some}
high-energy physics, perhaps string theory or an exotic field theory,
which gives rise to a 3-brane moving inside a bulk $d$-dimensional
spacetime. It is assumed that the SM, or some extension of it (which for
convenience we will continue to call the ``SM''), is
constrained to propagate within the 3-brane world-volume, and that
gravity, and perhaps some other degrees of freedom, are free to
propagate in the bulk. We then  
construct the most general consistent effective field theory 
describing the couplings between bulk gravity, the 3-brane fluctuations,
and the SM particles. By this means we can
 outpace the fundamental theorists who must
first tackle {\it how} a realistic 3-brane arises.

The organization of the paper is as follows. Section 2 sets up some
basic notation. Section 3 describes the couplings of the purely bosonic
degrees of freedom. The result of this section may appear rather obvious to
anyone familiar with the literature on D-branes, and no great
originality is claimed. Section 4 develops the formalism needed to
include chiral fermions, such as quarks and leptons, on the
3-brane. Section 5 explains how to gauge-fix reparametrization invariance of
the 3-brane description. Section 6 discusses the 
power-counting needed to implement renormalization in the effective
field theory program. Section 7 discusses the sense in which 
 the effective field theory is really a ``gauged chiral lagrangian''
 corresponding to the spontaneous breaking of higher-dimensional 
spacetime symmetries by the 3-brane ground state. In particular, our
treatment is
an adaptation of Volkov's general formalism for treating spontaneous
breaking of spacetime symmetries \cite{volkov}. 
Section 8 provides the conclusions.

The present formalism is not explicitly supersymmetric, but it is hoped
that the extension to supersymmetry can be accomplished by 
methods similar to those of Section 4. It is expected that this will tie in
closely with earlier work on low-energy effective theories describing the 
spontaneous (partial) breaking  of (higher) supersymmetry. See for
example refs. \cite{ssb1} and \cite{ssb2}. Ref. \cite{ssb2} gives a more
complete list of references on this topic.

The present paper (in particular Section 6) 
assumes an aquaintance with the methodology of 
effective field theory. Ref. \cite{georgi} provides a good introduction to the
basic concepts and techniques, in the relatively simple context of
 pion physics. Ref. \cite{donoghue} describes how to interpret general
relativity as a quantum effective field theory.

\section{Preliminaries}

\subsection{Fields, coordinates, and related notation}

We are interested in four-dimensional SM fields living on the
world-volume of a $3$-brane, which in turn is free to move  in a
gravitating bulk  spacetime of dimension, $d > 4$. 
The fields we consider will be the minimal set needed to realize this scenario.
(Adding non-minimal fields poses no extra problem.)   For simplicity  we
take the $3$-brane world-volume topology to be ${\bf R}^4$, and the bulk
topology to be either ${\bf R}^d$ or ${\bf R}^4 \times {\bf T}^{d-4}$,
where ${\bf T}^k$ denotes a $k$-torus.

The coordinates of the bulk spacetime will be denoted $X^M$. The bulk
coordinate indices are capital letters from
the middle of the Roman alphabet, $M, N, ... = 0,...,d-1$. 
We reserve the lower-case letters from the middle of the Roman alphabet
to refer to just the last $d-4$ of these indices, $m, n,... = 4,...,d-1$.
If one is only interested in 
bosonic fields (Section 3), the components of the bulk
metric, $G_{MN}(X)$, can  be considered as the
fundamental gravitational degrees of freedom.   Otherwise, the gravitational
degrees of freedom in the bulk are the components of the $d$-bein,
$E^A_{~M}(X)$ (that is, the $d$-dimensional vielbein). The
{\it local Lorentz} indices in the bulk are  capital letters from
the beginning of the Roman alphabet, $A, B, ... = 0,...,d-1$. The
$d$-bein is related to the bulk metric, $G_{MN}(X)$, by
\begin{eqnarray}
E_{~M}^A(X)~ \eta_{AB}~ E_{~N}^B(X) &=& G_{MN}(X), \nonumber \\
E_{~M}^A(X)~ G^{MN}(X)~ E_{~N}^B(X) &=& \eta^{AB},
\end{eqnarray}
where $\eta_{AB}$ is the $d$-dimensional Minkowski metric. 
It is useful to  subdivide the local Lorentz indices
into two subsets: the first four denoted by letters from the beginning
of the Greek alphabet, $\alpha, \beta,... = 0,...3$, while the remaining
indices are denoted by lower-case letters from the beginning of the
Roman alphabet, $a, b,... = 4,...,d-1$.

The coordinates intrinsic to the 3-brane will be denoted
$x^{\mu}$. The 3-brane coordinate indices are chosen from the middle of
the Greek alphabet, $\mu, \nu,... = 0,...,3$. The {\it bulk} coordinates
describing the position occupied by a point $x$ on the 3-brane, are
denoted $Y^M(x)$. They are dynamical fields.

 The last fields required are those of the SM, which are all
functions of $x$, since they live only on the 3-brane. They come
in three types: scalar fields, vector gauge fields, and left-handed Weyl
spinors, denoted $\phi(x), A_{\mu}(x), \psi_{L}(x)$ respectively. Any
right-handed spinor fields can be made left-handed by charge conjugation
in the usual manner. Spinor and internal indices are 
suppressed because it is entirely straightforward to
replace them whenever desired.

\subsection{The ``vacuum'' state}

The effective field theory will describe the small-amplitude, long-wavelength
 fluctuations of the
dynamical fields about the following state:
\begin{eqnarray}
E_{~M}^A(X) &=& \delta_{~M}^A, ~~ G_{MN}(X) = \eta_{MN}; \nonumber \\
Y^M(x) &=& \delta^M_{~\mu} x^{\mu}, \nonumber \\
\phi(x) &=& v.
\end{eqnarray}
That is, we expand about a Minkowski bulk spacetime, with the 3-brane
occupying the subspace spanned by the four $X^{\mu}$-axes, and with the
intrinsic 3-brane coordinates, $x^{\mu}$, agreeing with the bulk
coordinates, $X^{\mu}$. We also allow some of the scalar fields on the
3-brane to be non-zero, but constant over the 3-brane.

\section{The Bosonic Effective Field Theory}

This section describes the construction of the effective field theory
when fermionic fields are absent. The procedure for adding fermions
is treated in the next section.

\subsection{Effective theory of gravity in the bulk} 

In isolation, the
bulk gravitational fields are described by an action,
\begin{equation}
S_{bulk} =  \int d^dX~ {\rm det}(E) \{- \Lambda + 2M^{d-2} R + ...\},
\end{equation}
where the $d$-dimensional Einstein-Hilbert action has been explicitly
written, with $M$ being the $d$-dimensional Planck mass and $R$ the
$d$-dimensional curvature scalar, $\Lambda$ is a $d$-dimensional
cosmological constant term, and the
ellipsis is the series of higher dimensional geometric invariants with
coefficients given by powers of $1/M$, multiplied by order one (or less)
dimensionless couplings. The effective field theory philosophy and
technology for using this non-renormalizeable action is essentially the
same as for the usual $d = 4$ case, which has been discussed in detail
in ref. \cite{donoghue}. 

If we are considering the $d$-dimensional spacetime to be of the form
${\bf R}^4 \times {\bf T}^{d-4}$,  then gravity
becomes effectively four-dimensional at distances larger than the radii
of the $(d-4)$-torus, with an effective Planck constant, $M_{Pl}$,  given by
$M_{Pl}^2 = M^{d-2} V_{\bf T}$, where $V_{\bf T}$ is the volume of the
$(d-4)$-torus. See refs. \cite{1} \cite{2} for further discussion.

\subsection{The induced metric on the 3-brane}
 
The distance between two infinitesimally separated points on the 3-brane,
 $x$ and $x + dx$, is given by,  
\begin{eqnarray} 
ds^2 &=& G_{MN}(Y(x))~ dY^M ~dY^N \nonumber \\
&=& G_{MN}(Y(x)) ~\frac{\partial Y^M}{\partial x^{\mu}} d x^{\mu} 
~\frac{\partial Y^N}{\partial x^{\nu}} d x^{\nu},
\end{eqnarray}
from which we deduce that the induced metric on the 3-brane is
\begin{equation}
g_{\mu \nu}(x) = G_{MN}(Y(x)) ~\partial_{\mu} Y^M  ~\partial_{\nu} Y^N.
\end{equation}
 Given eq. (2), it is clear
  that $g_{\mu \nu}(x)$ will consist of small fluctuations about the
four-dimensional  Minkowski metric, $\eta_{\mu \nu}$.

\subsection{Effective field theory associated to the 3-brane}

Let us write the most general action for the bosonic fields associated to the
3-brane in the background of the bulk metric, $G_{MN}(X)$. 
The action must be invariant under general
$X$-coordinate transformations {\it as well as} $x$-coordinate transformations.
The requirement of the first of these invariances is clear since it is
the ``gauge invariance'' for the $d$-dimensional general relativistic
bulk gravity. The requirement of $x$-coordinate invariance follows
because the $x$-space is completely unphysical, just providing a
convenient means of parametrizing the 3-brane embedding, $Y(x)$. 

The book-keeping to enforce these two invariances is straightforward.
 We first have to determine how our fields transform under the two
types of coordinate transformations. $G_{MN}(X)$ is an
$X$-space 2-tensor and $x$-scalar. The action cannot depend directly on
all of $Y(x)$ because it makes reference to the origin of $X$-coordinate
space, which is unphysical 
(the usual statement that coordinates are not themselves
generally-covariant tensors), 
but the action can depend on  $\partial_{\mu} Y^M$, 
which is an $X$-vector and $x$-vector. $\phi(x)$ is obviously a
scalar of both spaces. $A_{\mu}(x)$ is an $x$-vector and an
$X$-scalar.\footnote{This point deserves some explanation. The gauge field
$A_{\mu}(x)$ describes the parallel transport between two infinitesmally
separated points on the 3-brane, $x$ and $x + dx$. The parallel
transport is as usual given by $1 + i A_{\mu}(x) dx^{\mu}$. Under an
$x$-coordinate transformation, $dx^{\mu}$ transforms covariantly,
and so $A_{\mu}$ must be taken to transform as a contravariant
vector. But under an $X$-coordinate transformation nothing
happens to the parallel transport, and so $A_{\mu}$ must be a scalar.} 
An important composite field is the induced 3-brane metric, $g_{\mu
  \nu}(x)$. From eq. (5), it is an $X$-scalar and an $x$-tensor. Using
$g_{\mu \nu}$ and $\partial_{\mu}$ we can construct covariant
$x$-derivatives in the standard way, and apply them to the various
tensors already discussed in order to generate further
tensors. Invariants can then be formed by contracting $x$-tensor indices using
$g_{\mu \nu}$ and its inverse, $g^{\mu \nu}$.

From these ingredients we can build the action,
\begin{eqnarray}
S^{3-brane}_{bosons} &=& \int d^4 x \sqrt{-g}\{- f^4 ~+~ \frac{g^{\mu \nu}}{2}
D_{\mu} \phi D_{\nu} \phi ~-~ V(\phi) \nonumber \\
&-& \frac{g^{\mu \nu} g^{\rho
\sigma}}{4} F_{\mu \rho} F_{\nu \sigma} + ...\},
\end{eqnarray}
where $F_{\mu \nu}(x)$ is the usual gauge field strength, and the ellipsis
includes higher-dimensional invariants one can build out of the fields
and covariant
derivatives, with dimensionful coefficients set by powers of
$1/f$ or $1/M$. (See Section 6.) 
Note that by locality, when $G_{MN}$ or its derivatives
appear in this action they must be evaluated on the 3-brane, at
$Y(x)$. The dominant terms, explicitly displayed in eq. (6),
  depend on $G_{MN}$ only through $g_{\mu \nu}$ and eq. (5), but higher
  invariants (in the ellipsis of eq. (6)) can certainly depend on
  $G_{MN}$ in a more general manner.

The leading  term of eq. (6)
 is a ``bare'' tension for the 3-brane, the mass scale $f$
being determined by the physics that gave rise to the 3-brane. This term
can be renormalized at tree- and loop-level by the vacuum energy of the
SM. The {\it renormalized} 3-brane tension dominates the interactions of bulk
gravity with the 3-brane when it is close to its ground state. This term
also contains kinetic energy for the $Y$ fields as will be discussed
in Section 6.

\section{Fermions on the 3-brane}

\subsection{The problem}

Recall how spin-$1/2$ fermions are ordinarily introduced in a
four-dimensional general relativistic context. (See ref. \cite{veltman}
for more details.) The fermions, $\psi(x)$, are regarded as
$x$-scalars, but as spinors of the local Lorentz group. The Lorentz
generators in the spinor representation are as usual given by,
 $\sigma_{(\alpha
  \beta)} \equiv \frac{1}{4}~ \gamma_{[\alpha} \gamma_{\beta]}$, where
  the $\gamma_{\alpha}$ are Dirac matrices satisfying,
\begin{equation}
\{\gamma_{\alpha}, \gamma_{\beta}\} = 2 \eta_{\alpha \beta}.
\end{equation}
The local
Lorentz group can formally be thought of as an {\it internal} $SO(3,1)$
gauge group. It gets related to spacetime through the vierbein,
$e^{\alpha}_{~\mu}$, an $x$-vector and local Lorentz vector which
satisfies,
\begin{eqnarray}
e^{\alpha}_{~\mu}(x) ~\eta_{\alpha \beta}~ 
e^{\beta}_{~\nu}(x) &=& g_{\mu \nu}(x), \\
e^{\alpha}_{~\mu}(x) ~g^{\mu \nu}(x)~
 e^{\beta}_{~\nu}(x) &=& \eta^{\alpha \beta}.
\end{eqnarray}

A covariant derivative for $\psi$ with respect to local Lorentz
transformations can be constructed in terms of the vierbein,
\begin{eqnarray}
D_{\mu} &=& \partial_{\mu} ~+~ \frac{1}{2} \omega_{\mu}^{\alpha \beta} ~
\sigma_{(\alpha \beta)},  \\
\omega_{\mu}^{\alpha \beta} &=& \frac{1}{2}~ g^{\rho \nu}~
e^{[\alpha}_{~~\rho} \partial_{[\mu} e^{\beta]}_{~~\nu]} \nonumber \\
&+& \frac{1}{4} ~g^{\rho \nu}~ g^{\tau \sigma}~ e^{[\alpha}_{~~\rho}
e^{\beta]}_{~~\tau}~ \partial_{[\sigma} e^{\gamma}_{~\nu]}~
e^{\delta}_{~\mu}~ \eta_{\gamma \delta}.
\end{eqnarray}
An $x$-coordinate invariant and local Lorentz invariant action then
follows,
\begin{equation}
S_{fermion} = \int d^4 x \sqrt{-g} \{\overline{\psi}~ i e^{\mu}_{~\alpha}
\gamma^{\alpha} D_{\mu} \psi + ...\}, 
\end{equation} 
where $e^{\mu}_{~\alpha}$ is the inverse of the vierbein, obtained from
$e^{\alpha}_{~\mu}$ by using $g^{\mu \nu}$ 
to raise the $x$-coordinate index and
$\eta_{\alpha \beta}$ to lower the local Lorentz index.

It therefore appears that incorporating fermions on the 3-brane will
require deriving the vierbein induced from the bulk $d$-bein and the
3-brane embedding. Indeed this is generally the case, and, as will be
seen below, the result is considerably more complicated than for the
induced metric, eq. (5). However, before embarking on this exercise it
is enlightening
 to see why we cannot generally get away with a simpler approach.

We can consider the four-dimensional local Lorentz group at $x$ on the
3-brane to be the subgroup of the $d$-dimensional local Lorentz group at
$Y(x)$ which acts non-trivially on the four-dimensional hyperplane
tangent to the 3-brane (spanned by $\partial_{\mu} Y^M~ E^A_{~M}$). It
follows that we get four-dimensional local Lorentz invariance by
demanding $d$-dimensional local Lorentz invariance. Just as in four
dimensions, we can construct $d$-dimensional gauge fields for local
Lorentz invariance, $\Omega^{AB}_M(X)$, in terms of  the $d$-bein. The
formula is exactly analogous to eq. (11). From this we get an induced
gauge field and covariant derivative on the 3-brane,
\begin{eqnarray}
\omega^{AB}_{\mu}(x) &\equiv& \partial_{\mu} Y^M~ \Omega_M^{AB}(Y(x)),
\nonumber \\
D_{\mu} &\equiv& \partial_{\mu} ~+~ \frac{1}{2} \omega_{\mu}^{AB}~
\Sigma_{(AB)},
\end{eqnarray}
where $\Sigma_{(AB)} \equiv \frac{1}{4}~ \Gamma_{[A}~ \Gamma_{B]}$ are
the $d$-dimensional Lorentz generators in spinor representation, and
$\Gamma_A$ are $d$-dimensional Dirac matrices satisfying,
\begin{equation}
\{ \Gamma_A, \Gamma_B \} = 2 ~\eta_{AB},
\end{equation}
We can therefore write an $x$-coordinate invariant and $d$-dimensional
local Lorentz invariant action,
\begin{equation}
S_{fermion} =  \int d^4 x \sqrt{-g}~ \{g^{\mu \nu} ~\partial_{\nu} Y^M~
E^A_{~M}~ \{\overline{\psi}~ i \Gamma_A
D_{\mu} \psi + ...\}. 
\end{equation}

Although eq. (15) is a consistent means of introducing fermions onto the
3-brane, it is not the most general way, and in particular does not
 give rise to four-dimensional chiral fermions. The reason is that
even if the  $\psi$ field appearing in eq. (15) is in an irreducible (perhaps
chiral) spinor representation  of the
$d$-dimensional local Lorentz group, it always 
corresponds to a {\it reducible} spinor
representation of the four-dimensional Lorentz subgroup. This reducible
representation contains equal numbers of left- and right-handed Weyl
doublets, as is familiar from dimensional reduction in Kaluza-Klein theory.

In superstring theory, fermions in the above reducible representation
naturally arise in the {\it simplest} D-brane configurations. 
They are some of the massless
modes of the open string that can attach to the D-brane. They are
related by supersymmetry to massless vector fields on the D-brane, and
so become  gauginos of the (highly-supersymmetric) 
low-energy gauge theory that lives on the D-brane. 

Clearly, in order to include two-component SM chiral fermions on the
3-brane we must adopt a different procedure. In fact we must explicitly
determine an induced vierbein on the 3-brane, as mentioned above. 
Then we can take our action to be given by, 
\begin{equation}
S^{3-brane}_{chiral-fermion} = 
\int d^4 x \sqrt{-g} \{\overline{\psi}_L ~i e^{\mu}_{~\alpha}
\sigma^{\alpha} D_{\mu}~ \psi_L + y \phi \psi_L \psi_L + {\rm h.c.} + ...\}, 
\end{equation}
where the $\sigma^{\alpha}$ are the usual $2 \times 2$ chiral Dirac matrices
for four-dimensional Minkowski space. The covariant derivative now
contains the local Lorentz gauge fields as in eqs. (10, ~11) as well as
gauge fields for internal gauge groups. Yukawa couplings to scalars are
also included.  
The ellipsis contains higher dimension interactions that
can be constructed with the help of the vierbein and covariant derivatives.
Of course we have the usual
requirement of cancellation of chiral gauge anomalies in order for our
effective theory to make sense at the quantum level.

\subsection{The induced vierbein}

The vierbein can conveniently be thought of as a means of
finding the components of the $x$-space differential, $d x^{\mu}$, in
(four-dimensional) local Lorentz coordinates, the result being just
$e^{\alpha}_{~\mu} d x^{\mu}$. Our strategy for obtaining the vierbein
from the 3-brane embedding is as follows. At each point $x$, we will
lift $d x^{\mu}$ to the corresponding infinitesimal $X$-space 
vector tangent to the 3-brane,
\begin{equation}
d Y^M = \partial_{\mu} Y^M d x^{\mu}.
\end{equation}
 Then we will perform a local $d$-dimensional Lorentz transformation 
at $Y(x)$, mapping the tangent hyperplane to lie
in the $\alpha = 0,...,3$ directions. In particular, $dY$ will be mapped
to an infinitesimal vector with non-zero components only in the $\alpha$
directions, $e^{\alpha}_{~\mu} d x^{\mu}$. The $e^{\alpha}_{~\mu}$ so
obtained will be proven to form a vierbein. This approach has
similarities with
the constructions of refs. \cite {bandos} \cite{ssb1} \cite{ssb2}. 
For a different approach to chiral fermions on branes see ref. \cite{holdom}.

The requisite Lorentz transformation, $R$, is determined as follows. Among the
$d$-dimensional Lorentz generators (in the vector representation), 
$J^{(AB)}$, the subgroup generated by
the $J^{(\alpha \beta)}$ and the $J^{(ab)}$ leaves invariant 
the subspace spanned by the $\alpha$ directions. We will drop these
generators and consider $R$ to be of the form,
\begin{equation}
R(x) = {\rm exp}(i \theta_{\alpha a}(x) J^{(\alpha a)}).
\end{equation}
The condition that the tangent hyperplane is mapped to lie in the
$\alpha$-directions is equivalent to requiring the Lorentz-transformed
tangent vectors to be orthogonal to the $a$-directions. That is,
\begin{equation}
R^a_{~B} ~E^B_{~M}(Y) ~\partial_{\mu} Y^M = 0,~ {\rm for
   ~all~} ~a, \mu.
\end{equation}
The $d$-bein has been used here to express $\partial_{\mu} Y$ 
in local Lorentz coordinates.

Eqs. (18) and (19) uniquely determine $R$, since they correspond to $4
\times  (d -4)$ equations for $4 \times (d-4)$ unknowns, $\theta_{\alpha
  a}(x)$, and we are expanding about field values eq. (2) for which there
is a unique solution, $\theta_{\alpha a} = 0$. Eq. (19) can  
be solved perturbatively to any desired order in the small fluctuations
about eq. (2). The precise algorithm for doing this is described in the
appendix. 

The  vierbein is then given by,
\begin{equation}
e^{\alpha}_{~\mu} \equiv R^{\alpha}_{~A} ~E^A_{~M}(Y) ~\partial_{\mu} Y^M.
\end{equation}
Let us prove that this 
indeed satisfies the properties of a vierbein, eqs. (8,~ 9). Eq. (20) implies,
\begin{eqnarray}
e^{\alpha}_{~\mu} ~\eta_{\alpha \beta}~ e^{\beta}_{~\nu} &=& 
R^{\alpha}_{~A} ~E^A_{~M} ~\partial_{\mu} Y^M  ~\eta_{\alpha \beta}~ 
R^{\beta}_{~B} ~E^B_{~N} ~\partial_{\nu} Y^N.
\end{eqnarray}
Now, by eq. (19) we can replace the sums over $\alpha, \beta = 0,...,3$ on
the right-hand side by sums from $0$ to $d-1$, since the sums from $4$ to
$d-1$ add nothing. Therefore,
\begin{eqnarray}
e^{\alpha}_{~\mu} ~\eta_{\alpha \beta}~ e^{\beta}_{~\nu} &=& 
R^{E}_{~A} ~E^A_{~M} ~\partial_{\mu} Y^M  ~\eta_{EF}~ 
R^{F}_{~B} ~E^B_{~N} ~\partial_{\nu} Y^N.
\end{eqnarray} 
The fact that $R$ is a $d$-dimensional Lorentz transformation implies
that,
\begin{equation} 
R^E_{~A} ~\eta_{EF} ~R^F_{~B} = \eta^{AB}.
\end{equation}
So eq. (22) simplifies to,
\begin{eqnarray}
e^{\alpha}_{~\mu} ~\eta_{\alpha \beta}~ e^{\beta}_{~\nu} &=& 
E^A_{~M} ~\eta_{AB} ~E^B_{~N} ~\partial_{\mu} Y^M 
 ~\partial_{\nu} Y^N \nonumber \\
&=& G_{MN} ~\partial_{\mu} Y^M  ~\partial_{\nu} Y^N \nonumber \\
&=& g_{\mu \nu},
\end{eqnarray}
where the second equality follows from eq. (1) and the third equality
from eq. (5).  Thus eq. (8) holds.
Regarding eq. (24) in matrix notation,
\begin{equation}
e^T \eta ~e = g,
\end{equation}
we can invert both sides, and then pre-multiply by $e$ and post-multiply
by $e^T$ to get eq. (9). 

We have found an induced
vierbein  which we can use to construct the action for
chiral fermions on the 3-brane, according to eq. (16). The full action
of our effective field theory is the sum of eqs. (3, ~6, ~16).

\section{Gauge-fixing the reparametrization invariance}

Our formalism up to this point has been explicitly 
invariant under general $x$-coordinate
transformations. This corresponds to a large reparametrization
invariance in our description of the 3-brane. If $Y^M(x)$ describes a
3-brane configuration and if $x'(x)$ is a general $x$-coordinate
transformation, then $Y^M(x'(x))$ describes an identical 3-brane
configuration. 

Fortunately, it is straightforward to eliminate this redundancy, leaving
us with only the physical number of 3-brane degrees of freedom. This is
done by imposing the gauge condition, 
\begin{equation}
Y^{\mu}(x) - x^{\mu} = 0, 
\end{equation}
while the $d-4$ fields, $Y^m(x)$, are physical and can fluctuate. 
For small fluctuations about
eq. (2), eq. (26) can always be solved. Note that this is a complete
gauge-fixing because if $Y$ satisfies eq. (26), then,
\begin{equation}
Y^{\mu}(x'(x)) - x^{\mu} = 0~~ {\rm if~ and~ only~ if~} ~x'(x) = x.
\end{equation}

In the quantum functional integral we need only integrate over $Y(x)$
which satisfy eq. (26). Furthermore, this gauge condition has a trivial 
ghost determinant, since if $Y$ satisfies eq. (26) and $x'(x) = x +
\xi(x)$ is an {\it infinitesimal} coordinate transformation,
\begin{eqnarray}
\frac{\delta}{\delta \xi^{\nu}(y)}_{|_{\xi = 0}} 
[Y^{\mu}(x'(x)) - x^{\mu}] &=& \partial_{\nu} Y^{\mu}(x) \delta^4(y - x)
\nonumber \\
&=& \delta_{~\nu}^{\mu} \delta^4(y-x),
\end{eqnarray}    
which is field independent.

\section{Power-counting, canonical fields, and renormalization}

The effective field theory construction 
described in the preceding sections admits and contains various types of
non-renormalizable interactions. We need to determine the
power-counting dimension, and thereby the relevance,
 of these interactions, by writing our theory in
terms of canonically normalized fields. 

The gravitational fields can be decomposed as usual as \cite{veltman}, 
\begin{eqnarray}
E_{~M}^A(X) &=& \delta_{~M}^A + \frac{H^{A}_{~M}(X)}{M^{d/2 - 1}}, \nonumber \\
G_{MN}(X) &=& \eta_{MN} + \frac{H_{MN} + H_{NM}}{M^{d/2 - 1}} + 
\frac{H_{ML} ~H_{~N}^L}{M^{d - 2}}, 
\end{eqnarray}
where $H^A_{~M}(X)$ is the canonical graviton field, and its indices have
been raised and lowered using the Minkowski metric. 
We must also canonically normalize the 3-brane coordinate fields,
$Y^m(x)$ (where we are assuming that we have eliminated the
reparametrization invariance according to eq. (26)). We see that a
kinetic term quadratic in $Y^m$ results from 
the expansion of the leading term of eq. (6) in powers of 
(derivatives of) $Y^m$. A canonically normalized set
of fields, ${\cal Z}^m(x)$, can be introduced by writing,
\begin{equation}
Y^m(x) \equiv \frac{{\cal Z}^m(x)}{f^2}.
\end{equation}

It is a troublesome but necessary feature of the presence of fermions
in the effective field theory (via eq. (16)), that dependence on the
gravitons and $Y^m$ is implicit in the $\theta_{\alpha a}$ angles that
determine $R$ and the vierbein through eqs. (18, ~19, ~20). To determine
 the interaction vertices in eq. (16) we have to determine these angles
from eq. (19) perturbatively in powers of (derivatives of) $H$ and
$Y$, as described in the appendix. Fortunately, for any process, 
computed to some fixed loop order,
only vertices with a limited number of $H$ and $Y$ will contribute.

We now consider the structure of our effective field theory for the 
three possible cases, (i) $f \sim M$, (ii) $f \ll
M$, (iii) $f \gg M$. 

(i) $f \sim M$: For power-counting purposes we can take all
higher-dimension interactions involving canonical fields to be of order
a  power of $1/M$, given by dimensional analysis. In order to make
sense of this non-renormalizable theory with an infinite number of
possible terms (in the ellipses of eqs. (3, ~6, ~16)) we must restrict
its domain of validity to momenta and field fluctuations (away from
eq. (2)) much smaller than $M$. For processes outside this domain, we
require a more fundamental description of quantum gravity and the
physics that gave rise to the 3-brane. The effective field theory
procedure in the domain of validity is to work to some fixed but
arbitrary order in $1/M$, say ${\cal O}(1/M^k)$, balanced by powers of
fields and momenta for the process under consideration. We then throw away all
terms in our effective lagrangian of higher order, leaving only a finite
number of interactions.

Now, if we only wish to do {\it classical} field theory, we can simply
use the truncated effective lagrangian. In the modern effective field
theory view, this is precisely the sense in which ordinary classical
general relativity, using only the Einstein-Hilbert action, 
is a valid approximation. However we can also do {\it quantum} effective
field theory. In computing Feynman diagrams we will encounter local
ultraviolet divergences which are formally of higher order than
$1/M^k$. We  can simply throw them away. The remaining divergences
(finite in number) will correspond precisely to the (counter-)terms we
have retained in our effective lagrangian, so renormalization can
proceed in the usual way.

(ii) $f \ll M$: For power-counting purposes, eq. (30) suggests that the
strength of non-renormalizable interactions involving only canonical
3-brane fields should be taken of order a power of $1/f$, the
scale fixed by the physics which gave rise to the 3-brane. Eq. (29) then
suggests that interactions involving extra gravitons, $H$, are further
suppressed by powers of $1/M$.
Now our effective field theory is valid for momenta and field
fluctuations much smaller than $f$. For processes beyond this domain, we
require a more fundamental description of the physics that gave rise to
the 3-brane. However, this may  not require a more fundamental
description of quantum gravity, the present general relativistic
description continuing to make sense for momenta all the way up to $M$.

The effective field theory procedure is now to do a double expansion. We
must work to some fixed but arbitrary order in $1/f$, balanced by powers
of 3-brane field fluctuations and momenta for the process under
consideration, and to some fixed order in
$1/M$, balanced by powers of the graviton field and momenta. For
example, if $f/M$ is small enough it may be a good approximation to work
to some non-trivial order in $1/f$, but to zeroth order in $1/M$. In this
approximation we are simply neglecting bulk gravity altogether, as we
frequently do in SM applications, but we are retaining the 3-brane
fluctations in the flat bulk spacetime. Once again, to any order in the
double expansion,
renormalization proceeds in the usual manner once the effective
lagrangian and ultraviolet divergences are truncated to the finite number
that are within the order to which we are working.

(iii) $f \gg M$: This is the case of a ``large'' 3-brane tension. In
this case it is quite unnatural to expect that the higher-dimension 
interactions involving 3-brane fields are suppressed by powers of $1/f$,
even if the interactions contain no explicit gravitons. The reason is
that gravity couples to everything and gravitational loops will dress
all possible interactions. We can therefore expect that any
higher-dimension interaction will naturally be of order powers of $1/M$,
unless protected by $X$-coordinate or $x$-coordinate invariance. That
is, for power counting purposes, we should first write our effective
lagrangian in terms of $X$-coordinate and $x$-coordinate invariants as
we have in eqs. (3, ~6, ~16). Then the naive coefficients of the various
higher-dimension invariants should be given by powers of $M$ determined
by dimensional analysis. The right powers of $1/f$ will then emerge when
the effective lagrangian is expanded in terms of the canonical fields.

While in principle any ultraviolet regulator can be used to regulate the
Feynman diagrams of the effective theory, it is of course preferable to
use a regularization that respects as many of the symmetries of the
theory as possible. The simplest procedure appears to be dimensional
regularization where one analytically continues the dimensionality of
both the 3-brane as well as that of the bulk spacetime. As is always the
case, this regularization does not respect chiral gauge invariance when
fermions are present on the 3-brane, but this nuisance is no more severe
than in ordinary four-dimensional field theories.

\section{The formalism of spontaneous symmetry-breaking}

The effective field theory developed above is a particular case of the chiral
lagrangian approach to spontaneous symmetry-breaking, and it is quite
useful to understand the deep sense in which this is so. Let us recall
the broad essence of this method. The chiral
lagrangian is a low-energy theory for the Nambu-Goldstone modes
associated to  spontaneously broken symmetries. If the full group of
dynamical symmetries is $G$, and the vacuum spontaneously breaks this
down to a subgroup $H$, then the Nambu-Goldstone modes transform under
all of $G$, but the transformations outside of $H$ are realized
non-linearly. The chiral lagrangian dynamics 
is tightly constrained to respect
the full $G$ symmetry. If there are other low-energy fields which are
not Nambu-Goldstone modes, but which transform linearly under $H$, they
are to be included in the chiral lagrangian, and coupled to the
Nambu-Goldstone modes so that the full $G$-invariance is
respected. Although initially  $G$ is taken to be a {\it global}
symmetry group, it can subsequently be weakly gauged in a
straightforward manner at the level of the chiral lagrangian. The beauty
of this method is that it separates the question of what the
low-energy {\it consequences} of spontaneous symmetry-breaking are from the
(frequently more difficult) question of what the {\it dynamical mechanism} for
the spontaneous symmetry-breaking is.

In general,
there are two types of symmetry that can be spontaneously broken, the
familiar case of internal symmetries and  the less familiar case of spacetime
symmetries. The general formalism for constructing the chiral lagrangian
in the former case was worked out in ref. \cite{ccwz}, while for the latter
case the formalism was provided in ref. \cite{volkov}. In this
paper, spacetime symmetry is spontaneously broken, and
this symmetry is  ``weakly gauged'' by gravity. Let us begin by
 turning off gravity, leaving $d$-dimensional Minkowski
spacetime. Formally, we set $E_{~M}^A = \delta_{~M}^A$. From eq. (2) we see
 that the 3-brane vacuum spontaneously breaks the $d$-dimensional
Poincare symmetry by picking out a four-dimensional hyperplane to
occupy. Specifically, this breaks the translations transverse to the
3-brane, generated by $P_m$, and the Lorentz transformations that change the
orientation of the 3-brane, generated by $J^{\alpha a}$. The corresponding
Nambu-Goldstone modes are the $Y^m(x)$ and $\theta_{\alpha a}(x)$ (see
subsection 4.2) 
respectively. Using these modes, our effective theory, given by eq. (6)
plus eq. (16), is invariant under
the full $d$-dimensional Poincare symmetry, the four-dimensional
Poincare subgroup being linearly realized and the remaining symmetry
transformations being non-linearly realized. It is quite remarkable that
the dynamics of this purely four-dimensional theory can respect
$d$-dimensional Poincare invariance! The magic comes from the special
couplings to the Nambu-Goto modes. In this non-gravitational limit our
effective theory is essentially an adaptation of the general formalism
of ref. \cite{volkov}

Note that the $\theta_{\alpha a}$ are not independent
degrees of freedom from the $Y^m$. 
This is a peculiarity of spacetime symmetry-breaking and can be traced
back to the fact that both translations and Lorentz transformations
share the same conserved current, the $d$-dimensional 
energy-momentum tensor, whereas in
the case of internal symmetries, each generator has its own conserved
current. In particular,
 the fact that the effective theory is invariant under the full
$d$-dimensional Poincare group implies that in addition to the
usual four-dimensional 
energy-momentum tensor we have extra conserved currents, $T^{\mu
  m}(x)$, 
\begin{equation}
\partial_{\mu} T^{\mu m} = 0.
\end{equation}

Finally, we can turn gravity back on, thinking of it as weakly gauging
the spontaneously broken $d$-dimensional Poincare group. We now become aware
of another peculiarity of the spacetime symmetry breaking. Since the
$P_m$ are among the broken generators, in principle 
they do not provide good quantum
numbers for labelling the bulk quanta, in the present case just the
gravitons. Of course, far away from the 3-brane it does seem sensible to
label bulk quanta by their $d$-dimensional momenta, but the interactions
arising from our effective field theory between
 these quanta and the 3-brane quanta {\it violate momentum conservation}
 in the $m = 4,...,d-1$ directions. This is not hard to
 understand intuitively. 
When the bulk quanta are soft compared to $f$, the scale setting
 the 3-brane tension,  the 3-brane vacuum state 
appears approximately as a rigid wall, extending infinitely in the 
$\vec{x}$-directions. 
The bulk quanta can lose(gain) transverse momentum
to(from) this infinitely massive wall. On closer inspection we see that
the wall is not perfectly rigid, so that impacts from bulk quanta can create 
distortions in the wall, parametrized by $Y^m(x)$, which can propagate
along the wall and can also excite the SM modes. These processes satisfy
only a {\it local} version of transverse momentum conservation, namely 
eq. (31). Note however that four-dimensional momenta
 are well-defined global charges which are conserved in all processes, 
since they correspond to unbroken symmetry generators.

\section{Conclusions}

This paper has described the minimal effective field theory formalism
needed to explore the low-energy implications of a 3-brane universe. It
appears relatively straightforward to generalize the present
framework in several directions, for example, adding non-minimal  bulk fields,
making the formalism supersymmetric, considering more complicated
spacetime topologies, or considering more than one brane embedded in the
bulk spacetime. 

The effective field theory
 formalism may help address the questions pursued in ref. \cite{peskin},
 regarding the transmission of
 supersymmetry-breaking between branes in the scenario of ref. \cite{horava}. 
 Even the supersymmetry-breaking mechanism need not be explicitly
 described, since
its consequences can be pursued via the chiral lagangian approach to
supersymmetry-breaking initiated in ref. \cite{akulov}.

\section*{Acknowledgments}
I am grateful to Jonathan Bagger for encouraging me to write up this
paper, and to Martin Schmaltz for discussions and for reading a
first draft.
This research was supported by the U.S. Department of Energy under grant
\#DE-FG02-94ER40818. 

\appendix

\section*{Appendix}

Here we derive a procedure for determining the
$\theta_{\alpha a}(x)$ that appear implicitly (via eqs. (18, ~19, ~20)) in the
fermion action, eq. (16), in powers of the graviton and 3-brane
fluctuations. In the vector representation of the $d$-dimensional Lorentz
group, $SO(d-1,1)$, we can write the generators, $J^{(AB)}$, $A < B$, in
the explicit form, 
\begin{equation}
J^{(AB)~C}_{~~~~~~~~D} = i ~\eta^{CE} (\delta^A_{~E}~ \delta^B_{~D} -
\delta^A_{~D}~ \delta^B_{~E}).
\end{equation}

Let us define a set of small quantities in which we can perturb, by writing,
\begin{equation}
E^B_{~M}(Y)~ \partial_{\mu} Y^M \equiv \delta^B_{~\mu} -
\epsilon^{B}_{~\mu}.
\end{equation}
Thus $\epsilon$ contains the small fluctuations around eq. (2). We will
do perturbation theory by formally expanding in powers of $\epsilon$,
\begin{eqnarray}
\theta_{\alpha a} &=&  \sum_{n = 0} \theta^{(n)}_{\alpha a} \nonumber \\ 
R &\equiv& e^{i \theta_{\alpha a} J^{(\alpha a)}} = \sum_{n = 0}
R^{(n)}.
\end{eqnarray}
Substituting eq. (33) into eq. (19)  gives, 
\begin{equation}
 ~R^a_{~\mu} = R^a_{~B} ~\epsilon^B_{~\mu}.
\end{equation}

Now, to zeroth order in $\epsilon$ we have the obvious solution,
$\theta^{(0)}_{\alpha a} = 0$, $R^{(0)} = I$. Higher order solutions can be
obtained iteratively using eq. (35). Suppose that we have already
determined $\theta^{(0)},...,\theta^{(n)}$. Then $\theta^{(n+1)}$ is
determined as follows. The $n+1$ order term of
eq. (35)  reads,
\begin{equation}
R^{(n+1)a}_{~~~~~~~\mu} = 
R^{(n)a}_{~~~~B}~ \epsilon^B_{~\mu}.
\end{equation}
Given the simple exponential series expansion for $R$ in terms of
$\theta$, it is obvious that $R^{(n)}$ is a computable ${\cal
  O}(\epsilon^n)$ polynomial
in $\theta^{(0)},...,\theta^{(n)}$. It follows that $\theta^{(n+1)}$
does not appear on the right-hand side of eq. (36). The left-hand side has a 
simple linear
 dependence on $\theta^{(n+1)}$ which can be expressed by writing, 
\begin{equation}
R^{(n+1)a}_{~~~~~~~\mu} = i 
~\theta^{(n+1)}_{\gamma c}~ J^{(\gamma
  c)a}_{~~~~~~\mu} ~+~  
R^{(n+1)a}_{~~~~~~~~\mu_{|_{\theta^{(n+1)}=0}}}, 
\end{equation}
where the second term of the right-hand side of eq. (37) 
is to be computed in terms
of $\theta^{(0)},...,\theta^{(n)}$, with $\theta^{(n+1)}$ set to zero. 
 By eq. (32), the first term of the right-hand side of eq. (37) 
 is simply $\theta_{\mu c}^{(n+1)}~ \eta^{ac}$. Therefore subsituting eq. (37)
 into eq. (36) yields, 
\begin{equation}
\theta_{\mu c}^{(n+1)} = 
\eta_{ac} (R^{(n)a}_{~~~~B}~ \epsilon^B_{~\mu} ~-~ 
R^{(n+1)a}_{~~~~~~~\mu_{|_{\theta^{(n+1)}=0}}}).
\end{equation}


\begin{thebibliography}{99}
\bibitem{rubakov} V. A. Rubakov and M. E. Shaposhnikov, Phys. Lett. B125
  (1983) 136.
\bibitem{1} N. Arkani-Hamed, S. Dimopoulos and G. Dvali, {\it The
    Hierarchy Problem and New Dimensions at a Millimeter},
  hep-ph/9803315.
\bibitem{2} I. Antoniadis, N. Arkani-Hamed, S. Dimopoulos and G. Dvali,
{\it New Dimensions at a Millimeter to a Fermi and Superstrings at a
  TeV}, hep-ph/9804398.
\bibitem{holdom} B. Holdom, Nucl. Phys. B233 (1984) 413.
\bibitem{ant1} I. Antoniadis, Phys. Lett. B246 (1990) 377.
\bibitem{ant2} I. Antoniadis, C. Munoz and M. Quiros, Nucl. Phys. B397
  (1993) 515.
\bibitem{horava} P. Horava and E. Witten, Nucl. Phys. B460 (1996) 506,
  hep-th/9510209;  E. Witten, Nucl. Phys. B471 (1996) 135,
  hep-th/9602070; P. Horava and E. Witten, Nucl. Phys. B475 (1996) 94,
  hep-th/9603142; P. Horava, Phys. Rev. D54 (1996) 7561, hep-th/9608019.
\bibitem{dvali} G. Dvali and M. Shifman, Phys. Lett. B396 (1997) 64,
  hep-th/9612128.
\bibitem{bandos} I. A. Bandos and W. Kummer, {\it P-branes, Poisson
    Sigma-Models and Embedding Approach to $(p+1)$-Dimensional Gravity},
  hep-th/9703099.
\bibitem{dimopoulos} I. Antoniadis, S. Dimopoulos and G. Dvali, {\it
    Millimetre-Range Forces in Superstring Theories with Weak-Scale
    Compactification}, Nucl. Phys. B516 (1998) 70, hep-ph/9710204. 
\bibitem{ovrut} A. Lukas, B. A. Ovrut, K. S. Stelle and D. Waldram, {\it
    The Universe as a Domain Wall}, hep-th/9803235.
\bibitem{dienes} K. R. Dienes, E. Dudas and T. Ghergetta, {\it Extra
    Spacetime Dimensions and Unification}, hep-ph/9803466.
\bibitem{tye} G. Shiu and S.-H. H. Tye, {\it TeV Scale Superstring and
  Extra Dimensions}, hep-th/9805157.
\bibitem{quiros} A. Pomarol and M. Quiros, {\it The Standard Model from
    Extra Dimensions}, hep-ph/9806263.
\bibitem{price} J. C. Long, H. W. Chan and J. C. Price, {\it
    Experimental Status of Gravitational-Strength Forces in the
    Sub-Centimeter Regime}, hep-ph/9805217.
\bibitem{peskin} E. A. Mirabelli and M. Peskin, {\it Transmission of
    Supersymmetry Breaking from a 4-Dimensional Boundary},
  hep-th/9712214.
\bibitem{polchinski} J. Polchinski, {\it TASI Lectures on D-Branes},
  hep-th/9611050.
\bibitem{giveon} A. Giveon and D. Kutasov, {\it Brane Dynamics and Gauge
    Theory}, hep-th/9802067.
\bibitem{volkov} D. V. Volkov, Sov. J. Particles and Nuclei 4 (1973) 3;
  also see the review by V. I. Ogievetsky, Proc. of X-th Winter School
  of Theoretical Physics in Karpacz, vol. 1, Wroclaw (1974) 227.
\bibitem{ssb1} J. Hughes, J. Liu and J. Polchinski, Phys. Lett. B180
  (1986) 370.
\bibitem{ssb2} J. A. Bagger, Nucl. Phys. Proc. Suppl. 52A (1997) 362,
 hep-th/9610022.
\bibitem{georgi} H. Georgi, ``Weak Interactions and Modern Particle
  Theory'', Benjamin/Cummings Publishing Company, Menlo Park  (1984).
\bibitem{donoghue} J. Donoghue, Phys. Rev. D50 (1994) 3874; {\it
    Introduction to the Effective Field Theory description of Gravity},
  gr-qc/9512024.
\bibitem{veltman} M. Veltman, in {\it Methods of Field Theory}, Les
  Houches (1975) 265.
\bibitem{ccwz} S. Coleman, J. Wess and B. Zumino, Phys. Rev. 177 (1969)
  2239; C. Callan, S. Coleman, J. Wess and B. Zumino, Phys. Rev. 177
  (1969) 2247.
\bibitem{akulov} D. V. Volkov and V. P. Akulov, JETP Lett. 16 (1972)
  438; see also J. Wess and J. Bagger, ``Supersymmetry and
  Supergravity'', Princeton University Press (1992).




\end{thebibliography}
\end{document}